\documentclass[aps,twocolumn,showpacs,preprintnumbers,nofootinbib,prd,superscriptaddress,groupedaddress,10pt]{revtex4-2}

\makeatletter
\def\l@subsubsection#1#2{}
\def\l@subsubsubsection#1#2{}
\makeatother

\usepackage{graphicx,amssymb,amsmath,amsthm,amsfonts,epsfig,epsf,fixmath}
\usepackage[usenames]{color}
\usepackage{epstopdf}
\usepackage{mathtools}
\usepackage{mathrsfs}
\usepackage{relsize}
\usepackage{placeins} 
\usepackage{booktabs} 
\usepackage{siunitx} 
\DeclareSIUnit \parsec {pc} 
\usepackage{aas_macros}
\usepackage{bm}
\usepackage{dcolumn}
\usepackage{latexsym}
\usepackage{rotating}
\usepackage{longtable} 

\usepackage{enumerate}
\usepackage{tensor,multirow}
\usepackage{url}
\usepackage[linktocpage]{hyperref}

\newcommand{\tn}{\textnormal}

\newcommand{\dd}{\mathrm{d}}

\newcommand{\totder}[2]{\frac{\mathrm{d} #1}{\mathrm{d} #2}}

\begin{document}
\title{Extreme Love in the SPA: constraining the tidal deformability of supermassive objects with extreme mass ratio inspirals and semi-analytical, frequency-domain waveforms}
\author{
Gabriel Andres Piovano$^1$,
Andrea Maselli$^{2,3}$,
Paolo Pani$^{1}$}

\affiliation{$^{1}$ Dipartimento di Fisica, ``Sapienza" Università di Roma \& Sezione INFN Roma1, Piazzale Aldo Moro 5, 
00185, Roma, Italy}
\affiliation{$^{2}$ Gran Sasso Science Institute (GSSI), I-67100 L’Aquila, Italy}
\affiliation{$^{3}$ INFN, Laboratori Nazionali del Gran Sasso, I-67100 Assergi, Italy}

\begin{abstract} 
We estimate the accuracy in the measurement of the tidal Love number of a supermassive compact object through the detection of an extreme mass ratio inspiral~(EMRI) by the future LISA mission.
A nonzero Love number would be a smoking gun for departures from the classical black hole prediction of General Relativity.
We find that an EMRI detection by LISA could set constraints on the tidal Love number of a spinning central object with dimensionless spin $\hat a=0.9$ ($\hat a=0.99$) which are
approximately four (six) orders of magnitude more stringent than what achievable with current ground-based detectors for stellar-mass binaries.
Our approach is based on the stationary phase approximation to obtain approximate but accurate semi-analytical EMRI waveforms in the frequency-domain, which greatly speeds up high-precision Fisher-information matrix computations. This approach can be easily extended to several other tests of gravity with EMRIs and to efficiently account for multiple deviations in the waveform at the same time.
\end{abstract}
\maketitle

\section{Introduction}
During a binary inspiral, the tidal interactions between two compact objects become increasingly more relevant. The gravitational field of each object produces a tidal field on its companion, deforming its shape and multipolar structure. This effect can be quantified in terms of tidal-induced multipole moments, more commonly known as the tidal Love numbers~(TLNs)~\cite{poisson2014gravity}.

A remarkable result in General Relativity~(GR) is that the TLNs of BHs are precisely zero. This was first demonstrated for nonrotating BHs~\cite{Damour_tidal,Binnington:2009bb,Damour:2009vw,Gurlebeck:2015xpa} then extended for slowly rotating BHs~\cite{Poisson:2014gka,Pani:2015hfa,Landry:2015zfa}, and more recently it has been proved for Kerr BHs\footnote{We refer here to the \emph{conservative} tidal response which is directly related to the TLNs. For a BH the dissipative response is nonzero and directly connected to the tidal heating~\cite{Hartle:1973zz,poisson2014gravity}, whose phenomenological consequences in our context have been recently studied in details~\cite{Hughes:2001jr,Maselli:2017cmm,Maselli:2018fay,Datta:2019epe,Maggio:2021uge,Sago:2021iku,Cardoso:2022fbq}.} without any approximations~\cite{LeTiec:2020bos,Chia:2020yla,LeTiec:2020spy}.
This is generically not the case for BHs in modified gravity and for dark ultracompact objects without a horizon~\cite{Cardoso:2017cfl}, such as boson stars~\cite{Cardoso:2017cfl,Sennett:2017etc,Mendes:2016vdr}, gravastars~\cite{Pani:2015tga,Cardoso:2017cfl,Uchikata:2016qku}, anisotropic stars~\cite{Raposo:2018rjn} and other simple exotic compact objects~\cite{Cardoso:2019rvt} with stiff equation of state at the surface~\cite{Cardoso:2017cfl}.
In some cases it was found that the TLNs vanish only logarithmically as a function of the compactness in the BH limit~\cite{Cardoso:2017cfl}, providing a “magnifying glass” for near-horizon physics~\cite{Maselli:2018fay,Datta:2021hvm}.

Beside posing an intriguing problem of “naturalness” in Einstein’s theory~\cite{Porto:2016zng} and being associated with special emerging symmetries~\cite{Hui:2020xxx,Charalambous:2021kcz,Charalambous:2021mea,Hui:2021vcv}, the precise cancellation of the TLNs for BHs in GR also provides an opportunity to test the prediction that all compact objects above a certain mass must be BHs: measuring a nonvanishing TLN  would provide a smoking gun for GR deviations or for the existence of new species of ultracompact massive objects.
The latter possibility is particularly relevant for supermassive objects which, in the standard paradigm, can only be BHs.

It has been recognized that, for extreme mass-ratio inspirals~(EMRIs), the TLNs of the central object affect the gravitational waveform at the leading order in the mass ratio~\cite{Pani:2019cyc}. This property was used to estimate very stringent constraints on the TLNs through an EMRI detection, as achievable by the future space mission LISA~\cite{LISA:2017pwj,Barausse:2020rsu,LISA:2022kgy} and also by third-generation detectors such as the Einstein Telescope~\cite{Maggiore:2019uih,Kalogera:2021bya,Barsanti:2021ydd}.
However, the estimates in Ref.~\cite{Pani:2019cyc} were based on a Newtonian computation and a hand-waiving GW dephasing argument which neglects correlations among different waveform parameters. The latter can jeopardize the detectability of a given effect even when the corresponding dephasing is significant~\cite{Piovano:2021iwv}.

The scope of this paper is to perform a proper estimate of the measurability of the TLNs in an EMRI signal. We shall focus on circular equatorial orbits, but include the spin of both binary components (henceforth the primary and the secondary). 
Beside being unavoidably present in the waveform, including the spin of the secondary in our context is also useful to understand whether a putatively small effect such as that induced by the tidal deformability of the primary can be confused by other small effects like those induced by the secondary spin.

We shall use the Fisher information matrix, which require to efficiently compute numerical derivatives of the waveform in terms of some of its parameters. For EMRIs, this task is highly delicate and time consuming~\cite{Maselli:2021men,Burke:2020vvk,Speri:2021psr,Piovano:2021iwv}.
To overcome known difficulties related with the inversion of the Fisher matrix and with the numerical derivatives, here we implement a semi-analytical approximation of the waveform using the stationary phase approximation~(SPA)~\cite{Hughes:2021exa,Damour:2000gg,Droz:1999qx}, which provides an accurate description in the frequency domain.
Although we apply this method to the estimate of the TLNs, we envisage that the same approach (with all its benefits) can be directly used for the many other tests of gravity with EMRIs~\cite{Maselli:2021men,Barausse:2020rsu,LISA:2022kgy,Babak:2017tow}.

Our main result is to confirm LISA's unique power in constraining the TLNs of a supermassive objects~\cite{Pani:2019cyc}. Although our projected bounds are, as expected, less optimistic than those naively derived in Ref.~\cite{Pani:2019cyc}, they remain remarkable: as detailed below we find that an EMRI detection with LISA at signal-to-noise ratio~(SNR) equal $30$ could constrain the TLNs of a highly-spinning supermassive object up to \emph{six orders of magnitude better} than what currently achievable with LIGO/Virgo for stellar-mass binaries~\cite{LIGOScientific:2018cki}.

We use $G=c=1$ units throughout and the notation follows that of Ref.~\cite{Piovano:2020zin}.

\section{Setup}\label{sec:setup}
Before providing the details of our model, it is useful to recall the general argument presented in Ref.~\cite{Pani:2019cyc}. Therein, it was recognized that, at leading post-Newtonian~(PN) order and in the small mass-ratio limit ($q\ll1$),
the tidal correction to the instantaneous GW phase reads 
\begin{equation}
    \phi_{\rm tidal}(f)\propto \frac{k_1}{q} v^5 \,,
\end{equation}
where $k_1$ is the (quadrupolar, electric) TLN of the primary, $f$ is the GW frequency, $v=(\pi M f)^{1/3}$, and $M$ is the mass of the primary. Thus, this correction enters at the same (adiabatic) order in the mass ratio as the ordinary radiation-reaction term, $\phi_{N}(f)\propto v^{-5}/q$, while being suppressed relative to the latter by a relative 5PN ($v^{10}$) factor.
If $k_1\gg q$, the tidal contribution is larger that the first-order correction due to the conservative part of the self force~\cite{Barack:2009ux,Poisson:2011nh}, which is instead suppressed by a factor ${\cal O}(q)$ relative to $\phi_{N}$.

This hand-waiving argument is based on a PN expansion, which is known to converge poorly in the extreme mass-ratio limit~\cite{Fujita:2011zk,Fujita:2012cm,Sago:2016xsp}. 
On the other hand, it is intriguing that the 5PN suppression of the tidal term might not be relevant for an EMRI, since most of the signal is accumulated at the innermost stable circular orbit~(ISCO), when $v={\cal O}(1)$ and the orbital distance $r={\cal O}(M)$.

With this motivation in mind, below we provide a more detailed model to incorporate tidal effects in EMRIs.

\subsection{A model for a Kerr-like deformable object}
The vacuum region outside a spinning object is not necessarily described by a Kerr geometry due to the absence of Birkhoff's theorem beyond spherical symmetry. However, in the BH limit, any deviation from the multipolar structure of a Kerr BH dies off sufficiently fast~\cite{Raposo:2018xkf} within GR or 
in modified theories of gravity whose effects are confined near the radius of the compact object\footnote{
For EMRIs, assuming that the central object is described by the Kerr metric is also well justified for gravity theories with higher curvature corrections to GR~\cite{Berti:2015itd}. In that case, the corrections to the metric are suppressed by powers of $l_P/M \ll 1$, where $l_P$ is the Planck length or the length scale of new physics~\cite{Maselli:2020zgv,Piovano:2020zin}.
}.
Explicit examples of this ``hair-conditioner theorem''~\cite{Raposo:2018xkf} within GR are given in 
Refs.~\cite{Pani:2015tga,Uchikata:2015yma,Uchikata:2016qku,Yagi:2015hda,Yagi:2015upa,Posada:2016xxx}, whereas examples in low-energy effective string theory were recently studied in the context of BH microstate geometries in Refs.~\cite{Bena:2020see,Bianchi:2020bxa,Bah:2021jno}.
In this regime, we assume that the background geometry of the primary is 
described by the Kerr metric (see e.g.~\cite{Maggio:2017ivp,Abedi:2016hgu} for similar models), given in Boyer-Lindquist coordinates by 
\begin{align}
ds^2=&-dt^2+\Sigma(\Delta^{-1}dr^2+d\theta^2)+(r^2+a^2)\sin^2\theta d\phi^2\nonumber\\
&+\frac{2Mr}{\Sigma}(a\sin^2\theta d\phi-dt)^2\ ,
\end{align}
where $\Delta=r^2-2Mr+a^2$, $\Sigma=r^2+a^2\cos^2\theta$, 
and $a$ is the spin parameter such that $\vert a\vert \leq M$. 
Without loss of generality, we consider the spin of the 
primary to be aligned with the $z$-axis, namely $a\geq0$. 
However, at variance with the standard BH picture, we will allow the object to be deformable when immersed in an external tidal field, in the sense that its TLNs are nonzero. 

Note that this model is \emph{conservative} since, besides including a nonzero TLN, the rest of the geometry is identical to that of a Kerr BH. In specific models of deformable supermassive objects one would generically expect also other deviations, such as tidal heating and deformed multipole moments (see~\cite{Cardoso:2019rvt} for a review).

\subsection{Orbital dynamics and radiation reaction effects}
We focus on circular, equatorial, and prograde orbits, for which the initial 
angular momentum $L_z$ is positive and parallel to the $z$-axis.
To avoid the complications induced by spin precession, we assume that also the secondary spin is (anti)-aligned with the primary spin.

The EMRI orbital evolution is driven by adiabatic 
Teukolsky fluxes \cite{Huerta:2011zi}, 
including linear corrections due to the secondary spin.  
The radiation reaction equations for the evolution of 
the orbital parameters are expanded in the mass ratio, 
and include the contributions due to the TLN of the primary. As explained below, the latter are included in a PN fashion. 
Although this hybrid model combines elements of BH perturbation 
theory with PN terms, it allows describing the EMRI dynamics in the strong field regime near the primary, at variance with a fully PN 
description of the orbital dynamics that instead breaks down near the ISCO.

The orbital motion of a spinning point particle in Kerr 
spacetime features two integrals of motion:
the normalized energy $\tilde E = E/\mu$ and angular 
momentum $\tilde J_z = J_z/(\mu M)$~\cite{Ehlers:1977}, where $\mu=q M\ll M$ is the secondary mass. 
To characterize the intrinsic angular momentum 
of the secondary, we introduce the dimensionless 
parameter
\begin{equation}
 \sigma = \frac{S}{\mu M} = \chi q \ , \label{def:sigma}
\end{equation}
where $\chi= S/\mu^2$ is the reduced spin of the secondary. 
For EMRIs, $|\chi|\ll 1/q$, which implies $|\sigma|\ll1$. 
This allows us to expand both $\tilde{E}$ and $\tilde{J}_z$ in 
terms of the spin parameter, considering linear corrections 
only, 
\begin{align}
\tilde E &= \tilde E^0 + \sigma \tilde E^1\quad \ , \quad \tilde J_z = \tilde J_z^0 + \sigma \tilde J_z^1 \,.
\end{align}
The explicit expressions of $\tilde E$ and $\tilde J_z$ are given in~\cite{Piovano:2021iwv}. We add to the binding energy $\tilde E$ the PN contribution $\tilde E_{\tn{TLN}}$ due to the TLN of the primary ($k_1$) in the point-particle limit~\cite{Abdelsalhin:2018reg,Pani:2019cyc} \footnote{Hereafter hatted quantities refer to dimensionless variables 
normalized to the primary mass, e.g $\hat r = r/ M$, $\hat a = a/ M$, $\widehat \Omega = M \Omega$, and so on.}
\begin{equation}
\tilde E_{\tn{TLN}} \simeq -\tilde E_{\tn{PN}}\bigg(\frac{6}{\hat r^5}+\frac{88}{3}\frac{1}{\hat r^6}\bigg)k_1\ ,\label{Ebinding}
\end{equation}
with $\tilde E_{\tn{PN}} = -q/(2  \hat r)$ being the leading-order binding energy in the PN expansion. In the above expression, we included both 5PN and 6PN tidal terms~\cite{Vines:2011ud}. Note that the secondary TLN ($k_2$) would contribute 
Eq.~\eqref{Ebinding} with terms scaling as $q^4$~\cite{Pani:2019cyc}, 
thus being largely subdominant for the EMRI case. 
The orbital frequency 
$\widehat \Omega$ is given by
\begin{equation}
\widehat\Omega(\hat r) = \widehat \Omega^0(\hat r) + \sigma \widehat \Omega^1(\hat r)\ ,
\end{equation}
where $\widehat{\Omega}^0(\hat r)=1/(\hat a\pm\hat r^{3/2})$ is the
Keplerian frequency for a nonspinning particle, and
\begin{equation}
 \widehat\Omega^1(\hat r)= -\frac{3}{2}\frac{\sqrt{\hat r}\mp \hat a}{\sqrt{\hat r}(\hat r^{3/2}\pm a)^2}\,.
\end{equation}
Once the orbital radius $\hat r$ and the parameters $\hat a$ and $\sigma$ are specified, the orbital dynamics is completely determined by $\tilde E, \tilde J_z $, and $\widehat\Omega$.

At the adiabatic level, the rate of change of the constants 
of motion $\tilde E$ and $\tilde J_z$ is balanced by the 
emitted GW fluxes, 
in which post-adiabatic corrections induced by the secondary 
spin are included as described in Ref.~\cite{Piovano:2021iwv}. 
These balance laws hold at first order in 
$\sigma$ for a spinning particle~\cite{Akcay:2019bvk}. 
The energy 
fluxes can also be expanded in $\sigma$ at fixed spins $\hat a$ and orbital radius $\hat r$~\cite{Piovano:2021iwv}:
\begin{equation}
\mathcal{F}(\hat r,\widehat\Omega) = \mathcal{F} ^0(\hat r,\widehat\Omega^0)+ \sigma \mathcal{F}^1(\hat r,\widehat\Omega^0,\widehat\Omega^1)+\mathcal{F}_{\rm TLN}(\hat r) \, ,
\end{equation}
where
\begin{align}
\mathcal{F} ^0+ \sigma \mathcal{F}^1
&= \frac{1}{q}\Bigg[ \bigg( \frac{\dd \tilde E}{ \dd \hat{t}} \bigg)^{\!\!H}_{\!\!\text{GW}} + \bigg( \frac{\dd 
\tilde E}{ \dd \hat{t}} \bigg)^{\!\!\infty}_{\!\!\text{GW}} \Bigg] \ , \label{TeuFluxes}
\end{align}
with $\left(\dd \tilde E/ \dd \hat t \right)^{\!\!H,\infty}_{\!\!\text{GW}}$ being the energy 
flux across the horizon and at infinity, respectively, as computed solving Teukolsky's equations.
The tidal contribution to the flux reads
\begin{equation}
\mathcal{F}_{\tn{TLN}}(\hat r) = \frac{128}{5}\frac{k_1}{\hat r^{10}}\Big(1-\frac{22}{21}\frac{1}{\hat r}\Big)\ , \label{fluxtidal}
\end{equation}
where again we have included both 5PN and 6PN corrections. 
Equation~\eqref{fluxtidal} shows that the TLN of the 
primary contributes to the GW fluxes at the leading, 
\textit{adiabatic}, order in $q$~\cite{Pani:2019cyc}. 

By defining
\begin{equation}
\mathcal{G}(\hat r,\widehat\Omega) \coloneqq \bigg( \totder{\tilde E}{\hat{r}}\bigg)^{\!-1} \mathcal{F}(\hat r,\widehat\Omega)\ ,
\end{equation}
then, at first order in the mass ratio,
\begin{align}
&\mathcal{G}(\hat r,\widehat\Omega) = \mathcal{G} ^0(\hat r,\widehat\Omega^0)  + \sigma \mathcal{G}^1(\hat r,\widehat\Omega^0,\widehat\Omega^1) + \mathcal{G}_{\mathrm{TLN}} \, ,\\
&\mathcal{G} ^0 =\bigg( \totder{\tilde E^0}{\hat{r}}\bigg)^{\!-1} \big(\mathcal{F}^0+\mathcal{F}_{\tn{TLN}}\big) \, ,\\
&\mathcal{G} ^1 =\bigg( \totder{\tilde E^0}{\hat r}\bigg)^{\!-1} \mathcal{F}^1 - \bigg( \totder{\tilde E^0}{\hat r}\bigg)^{\!-2} \bigg(\totder{\tilde E^1}{\hat{r}}\bigg)\big(\mathcal{F}^0+\mathcal{F}_{\rm TLN}\big)\,, \\
&\mathcal{G}_{\mathrm{TLN}} = - \bigg( \totder{\tilde E^0}{\hat r}\bigg)^{\!-2} \bigg(\totder{\tilde E_{\mathrm{TLN}}}{\hat{r}}\bigg)\big(\mathcal{F}^0+\mathcal{F}_{\rm TLN}\big)\,,
\end{align}
which yield for the time evolution of the orbital radius
\begin{equation}
\totder{\hat{r}}{\hat t} = - q\big[\mathcal{G} ^0(\hat r,\widehat\Omega^0)  + \sigma \mathcal{G}^1(\hat r,\widehat\Omega^0,\widehat\Omega^1)\big] \label{eq:radiusevol} \,.
\end{equation}
Likewise, at first order in $\sigma$ the orbital phase is given by
\begin{equation}
\totder{\phi}{\hat t} =\widehat{\Omega}^0(\hat r) + \sigma \widehat \Omega^1(\hat r) \,. \label{eq:phaseevol}
\end{equation}
Solving Eqs.~\eqref{eq:radiusevol} and \eqref{eq:phaseevol} and linearizing them in $\sigma$ yields the time evolution of 
$\hat r(\hat t)$ and $\phi(\hat t)$, 
which provide the basic ingredients to compute 
the GW signal emitted by the binary.
We compute the Teukolsky fluxes~\eqref{TeuFluxes} using the same setup and procedure detailed in Refs.~\cite{Piovano:2020ooe,Piovano:2020zin,Piovano:2021iwv}. Likewise, the time evolution of $\hat r(\hat t)$ and $\phi(\hat t)$ is performed as detailed in Ref.~\cite{Piovano:2021iwv}.

\subsection{Time-domain waveform}\label{sec:timewaveforms}
We use the quadrupole approximation for the GW strain~\cite{Huerta:2011zi}:
\begin{align}
h_\alpha(t)&=\frac{2\mu}{D} \widehat\Omega(t)^{2/3}\Big[A^+_\alpha(t)\cos(2\phi(t)+2\phi_0)+ \nonumber \\
&+A^\times_\alpha(t) \sin(2\phi(t)+2\phi_0)\Big] \, , \label{eq:gwsignal}
\end{align}
where $\alpha=I,II$ identifies two independent Michelson-like detectors that constitute LISA's response~\cite{Gourgoulhon:2019iyu},
\begin{align}
A^+_\alpha(t)&=(1+\cos^2\vartheta)F^+_\alpha(t)\,, \\
A^\times_\alpha(t)&=-2\cos\vartheta  F^\times_\alpha(t)  \, ,
\end{align}
where $\phi_0$ is the initial orbital phase, $D$ is 
the source's luminosity distance from the detector,
and $(\vartheta,\varphi)$ identify the direction, in 
Boyer-Lindquist coordinates, of the latter 
in a reference frame centered at the source. The 
antenna pattern functions $F^+_\alpha(t)$ and 
$F^\times_\alpha(t)$ depend on the angles 
$(\vartheta_S, \varphi_S)$ and $(\vartheta_K, \varphi_K)$ that provide the direction of the source and of the orbital angular momentum~\cite{Huerta:2011kt} in a heliocentric reference frame attached with the ecliptic\footnote{For equatorial orbits, $(\vartheta_K, \varphi_K)$ coincide with the direction of the primary spin.}~\cite{Barack:2006pq}.
The polar angle $\vartheta$ 
can be recast in terms of $(\vartheta_S, \varphi_S)$ and $(\vartheta_K, \varphi_K)$ as
\begin{equation}
\cos \vartheta = \cos \vartheta_S \cos \vartheta_K + \sin \vartheta_S \sin \vartheta_K \cos(\varphi_S - \varphi_K)\ .
\end{equation}
It is convenient to rewrite Eq.~\eqref{eq:gwsignal} in a 
more compact form:
\begin{align}
h_\alpha(t)&=\frac{2\mu}{D} \widehat\Omega(t)^{2/3}\mathcal A_\alpha(t) \cos(\Phi_\alpha(t)) \, ,  \label{eq:gwsignal2}\\
\Phi_\alpha(t) &= 2\phi(t) +2 \phi_0 +\phi^{\textup{sh}}_\alpha(t) \, , \\
\phi^{\textup{sh}}_\alpha(t) &= \arctan\bigg(-\frac{A^\times_\alpha(t)}{A^+_\alpha(t)}\bigg) \, , \\
\mathcal A_\alpha(t) & = \sqrt{(A^+_\alpha(t))^2+(A^\times_\alpha(t))^2} \, .
\end{align}

Finally, we include the effect of the Doppler modulation 
induced by the LISA orbital motion, by introducing 
a shift in the GW phase:
\begin{align}
&\Phi_\alpha(t) \to \Phi_\alpha(t) + \phi^{\textup{Dop}}(t) \, , \\
& \phi^{\textup{Dop}}(t) = 2\Omega(t)R\sin \vartheta_S \cos[2\pi (t/T_{{\rm LISA}})-\varphi_S] \ ,
\end{align}
where $R= 1{\rm AU}$ and $T_{{\rm LISA}}=1\,{\rm yr}$ is LISA's orbital period~\cite{Huerta:2011kt}.
\subsection{Frequency-domain waveform in the SPA}\label{sec:freqwaveforms}
We employ the SPA to obtain 
an approximate but accurate semi-analytical representation of 
the waveform templates in the frequency domain~\cite{Hughes:2021exa,Damour:2000gg,Droz:1999qx}. The Fourier transform of our time-domain waveform~\eqref{eq:gwsignal2} is given as:
\begin{equation}
\tilde h_\alpha(f) = \frac{\mu}{D}\!\int \limits_{-\infty}^\infty \!\!\!\mathrm{d}t \, \widehat\Omega(t)^{2/3}\mathcal A_\alpha(t) e^{-2\pi i f t }\big(e^{i\Phi_\alpha(t)}+e^{-i\Phi_\alpha(t)}\big) \ , \label{eq:FourierTransform}
\end{equation}
and we assume that $\Phi_\alpha$ is strictly monotonic in time, i.e. $\dot\Phi_\alpha(t) >0 $. We can rewrite Eq.~\eqref{eq:FourierTransform} as
\begin{align}
\tilde h_\alpha(f) &= \tilde h^+_\alpha(f) + \tilde h^-_\alpha(f) \ , \\
\tilde h^-_\alpha(f) &= \frac{\mu}{D} \!\int \limits_{-\infty}^\infty\!\!\!\mathrm{d}t \, \widehat\Omega(t)^{2/3}\mathcal A_\alpha(t) e^{-i(2\pi f t - \Phi_\alpha(t)) } \ , \\
\tilde h^+_\alpha(f) &= \frac{\mu}{D}\! \int \limits_{-\infty}^\infty\!\!\!\mathrm{d}t \, \widehat\Omega(t)^{2/3}\mathcal A_\alpha(t) e^{-i(2\pi f t + \Phi_\alpha(t)) } \ .
\end{align}
It is sufficient to compute the Fourier transform only for positive frequencies $f$, since our chirp signal is real. The integral $ \tilde h_\alpha(f)$ rapidly oscillates, and the contributions due to the complex exponential cancel out  except near the times interval $\tilde t$ where the Fourier phase $\Psi_\alpha \equiv 2 \pi f t  - \Phi_\alpha(t)$ is stationary:
\begin{equation}
\left. \totder{\Psi_\alpha}{t} \right \rvert_{t = \tilde t}  = 0 \implies 2\pi f  = \dot \Phi_\alpha (\tilde t) \ .
\end{equation}
In this case, $\tilde h^+_\alpha$ is negligible~\cite{Damour:2000gg} , thus $\tilde h_\alpha \approx \tilde h^-_\alpha$.
It is possible then to expand in Taylor series $ \Psi_\alpha$ near $\tilde t$:
\begin{equation}
\Psi_\alpha(t) = \Psi_\alpha(\tilde t) + \left.\frac{1}{2}\frac{\mathrm d^2 \Psi_\alpha} {\mathrm d t^2}\right \rvert_{t = \tilde t}(t- \tilde t )^2  +o\big((t- \tilde t )^3\big) \ .
\end{equation}
By plugging the above expansion in $\tilde h^-_\alpha(f)$, we obtain the following approximation of $\tilde h_\alpha(f)$:
\begin{equation}
\tilde h _\alpha (f) \simeq \frac{\mu}{D}e^{-i(2\pi f \tilde t - \Phi_\alpha(\tilde t))}  \!\! \int \limits_{-\infty}^\infty \!\! \!\mathrm{d}t \,\widehat\Omega(t)^{2/3}\mathcal A_\alpha(t) e^{-i\frac{1}{2}\ddot \Phi_\alpha (t -\tilde t)^2} \ .
\end{equation}
Before proceeding, we notice that  $\Phi_\alpha$ includes the terms $\dot\phi^{\textup{sh}}(t)$ and $\dot\phi^{\textup{Dop}}(t)$, which are suppressed  by a factor $2\pi/(\Omega(t) T_{\rm LISA})\ll1$. Thus, we can safely neglect these terms, approximating $\dot \Phi_\alpha(t) \approx 2\Omega(t)$. 
Further assuming that $\mathcal A_\alpha(t) $ is slowly varying with time, we can write (after a change of variables)
\begin{equation}
\tilde h _\alpha (f) \simeq \frac{\mu}{D}e^{-i(2\pi f \tilde t - \Phi_\alpha(\tilde t))}   (\pi M f)^{2/3}\mathcal A_\alpha(\tilde t) \!\!\int \limits_{-\infty}^\infty \!\! \!\mathrm{d}s \,e^{-i\dot \Omega(\tilde t)s^2} \ .
\end{equation}
The integral in the previous expression can be computed by standard techniques, leading to the SPA for the signal~\eqref{eq:gwsignal2}
\begin{align}
\tilde h_\alpha(f) &=\frac{\mu}{D} (\pi f M)^{2/3}\mathcal A_\alpha[\tilde t(f)] \sqrt{\frac{\pi}{|\dot \Omega(\tilde t(f))|}}e^{-i \tilde\Phi_\alpha[\tilde t(f)]} \, ,  \label{eq:gwSPA}\\
\tilde\Phi_\alpha[\tilde t(f)]&= 2 \pi f (\tilde t(f)+t_0) - 2(\phi(\tilde t(f)) + \phi_0) +\nonumber\\ 
&-\phi^{\textup{Dop}}(\tilde t(f)) -\phi^{\textup{sh}}_\alpha[\tilde t(f)] \pm \pi/4 \, .\label{eq:SPAphase}
\end{align}

The sign in Eq.~\eqref{eq:SPAphase} is fixed by the sign of the {frequency sweep} $\dot \Omega$, given by
\begin{equation}
\dot \Omega = \totder{\hat r}{t}\totder{\Omega}{\hat r}\ ,\label{sweep}
\end{equation}
while $\tilde t(f)$ is the time at which the equation
\begin{equation}
\Omega(t) = \pi f
\end{equation}
holds for any given Fourier frequency $f$. 
The SPA is accurate as long as the amplitude $\mathcal A(t)$ and orbital frequency $\Omega (t)$ are slowly varying:
\begin{align}
\left \lvert\frac{1}{\mathcal A_\alpha(t)}\totder{\mathcal A_\alpha(t)}{t}\right \rvert \sim \frac{\mathcal O(10)}{T_{\rm LISA}} \ll |\Omega(t)|\ \ ,\ \left\lvert\frac{\dot\Omega(t)}{\Omega(t)^2}\right\rvert \ll 1\ .
\end{align}
The first condition is always satisfied since for a typical EMRI $\Omega(t) T_{\rm LISA}\gg{\cal O}(10)$, while we have verified that the second criterion is met for all the binary configurations we analysed. Moreover, 
the SPA requires $\Omega(t)$ to be strictly monotonic during 
the orbital evolution. We have checked that this condition 
is also satisfied in our case (whereas it is not necessarily the case for more general orbits).
As a final remark, we note that the frequency-domain waveform 
is known fully analytically except for the orbital phase 
$\phi(t)$, the time $\tilde t(f)$, and  the frequency sweep $\dot \Omega (t)$,  which have implicit and explicit dependence on the parameters, and needs to be computed numerically.

\section{Accurate Fisher matrix analysis for EMRI waveforms}\label{sec:fisher}

The GW signal emitted by a circular, equatorial EMRI with a spinning secondary, 
moving on the equatorial plane with spin (anti)aligned to the $z$-axis, and including the tidal deformability of the primary, is completely specified by twelve parameters $\vec y =\{\vec y_\textnormal{I},\vec y_\textnormal{E}\}$:
(i) six intrinsic parameters $\vec y_\textnormal{I}=(\ln\mu, \ln M , \hat a,\chi, t_0, k_1)$ and (ii) six extrinsic parameters $\vec{y}_\textnormal{E}=(\phi_0,\vartheta_S, \varphi_S, \vartheta_K, \varphi_K, \ln D$). We remind that $(M,\mu)$ are the mass components with $q=\mu/M\ll1$, $(\hat a,\chi)$ are the primary and secondary spin parameters, $k_1$ is the dimensionless TLN of the primary, $(\phi_0,\hat t_0)$ define the binary initial phase and starting time, and $D$ is the source luminosity distance. The four angles $(\vartheta_S,\varphi_S)$ and $(\vartheta_K,\varphi_K)$ correspond to the colatitude and the azimuth of the source sky position and of the orbital angular momentum, respectively~\cite{Barack:2006pq}.
Since the orbit is circular and equatorial, the orbital angular momentum has no precession around the primary spin, and all angular momenta are parallel to each other.

In the limit of large SNR, the errors on the source parameters inferred by a given EMRI observation can be determined using the Fisher information matrix:
\begin{equation}
\Gamma_{ij} =\sum_{\alpha=I,II}^{} \left(\frac{d \tilde h_\alpha}{dy^i}\middle| \frac{d \tilde h_\alpha}{d y^j}\right)_{\vec y = \vec y_0}\ ,
 \end{equation}
where $\vec y_0$ corresponds to the true set of binary parameters, and we have introduced the noise-weighted scalar product between two waveforms $p_\alpha$ and $q_\alpha$ in the frequency domain: 
\begin{equation}
(p_\alpha|q_\alpha) = 2 \int_{f_{\rm min}}^{f_{\rm max}}\frac{df}{S_n(f)}
[\tilde p^*_\alpha(f)\tilde q_\alpha(f)+\tilde p_\alpha(f)\tilde q^*_\alpha(f)]\ , \label{innerproduct}
\end{equation}
where $S_n(f)$ corresponds to the noise 
spectral density of the detector, and the star 
identifies complex conjugation. 
The scalar product was computed using the Simpson's integration 
method.
In our computations we choose $f_{\rm max}$ and $f_{\rm min}$ as
\begin{align}
f_{\rm min}&=\frac{2}{2\pi}\frac{1}{M}\Big[\widehat{\Omega}^0(\hat r_0) + \sigma \widehat \Omega^1(\hat r_0)\Big]\ , \\
f_{\rm max}&=\frac{2}{2\pi}\frac{1}{M}\Big[\widehat{\Omega}^0(\hat r_{\rm ISCO}) + \sigma \widehat \Omega^1(\hat r_{{\rm ISCO}})\Big]\ ,\label{deffmax}
\end{align}
where $\hat r_{\rm ISCO}$ is location of the ISCO for a \emph{nonspinning} test particle around a spinning central object, and $\hat r_0$ is the initial orbital 
radius.
The waveform scalar product allows us to define the optimal 
SNR for a given signal $h$:
\begin{equation}
{\rm SNR} = (h|h)^{1/2}\,,
\end{equation}
which scales linearly with the inverse of the 
luminosity distance. 
The inverse of $\Gamma_{ij}$ is the covariance 
matrix $\Sigma_{ij}$, whose diagonal elements correspond 
to the statistical uncertainties of the waveform parameters,
\begin{equation}
\sigma^2_{x_i}=(\Gamma^{-1})_{ii}\ ,\label{fishererrors}
\end{equation}
whereas the off-diagonal elements correspond to the correlation coefficients.
In the large-SNR limit the covariance matrix scales 
inversely with the SNR. For a given set of parameters, 
it is therefore straightforward to rescale the errors by 
varying the luminosity distance $D$, and hence the SNR.

In addition to the standard deviations on the twelve parameters 
defined above, we also analyze the error box on the solid angle 
spanned by the unit vector associated to $(\vartheta_{S},\varphi_{S})$ and $(\vartheta_{K},\varphi_{K})$:
\begin{equation}
 \Delta\Omega_{i}=2\pi |\sin\vartheta_{i}|\sqrt{\sigma^2_{\vartheta_{i}}\sigma^2_{\varphi_{i}}
 -\Sigma^2_{\vartheta_{i}\varphi_{i}}}\ .
\end{equation}
where $i=(S,K)$.

As discussed in \cite{Piovano:2021iwv}, the inclusion of the 
secondary spin can severely deteriorate the accuracy 
with which the other intrinsic parameters are recovered. 
For this reason, we consider three alternatives scenarios 
in our data analysis: (i) the secondary spin $\chi$ is an unbounded parameter, (ii) a suitable prior is applied to $\chi$, 
(iii) the secondary spin is integrated out from the posterior distribution, by simply removing the corresponding row and column from the Fisher matrix.

For case (ii) we 
assume a wide prior given by a Gaussian probability distribution 
with standard deviation $\sigma_0 = 1$ and zero mean. In this 
configuration the errors on the source parameters are given by 
\begin{equation}
\sigma^2_{x_i}=\Sigma_{ii} = [(\Gamma+\Gamma_0)^{-1}]_{ii}\ ,
\end{equation}
where $(\Gamma_0)_{ij}=1/\sigma_0\delta_{i\chi}\delta_{\chi j}$ is the Fisher matrix corresponding to the prior 
distribution~\cite{Piovano:2021iwv}. \\

We have computed the numerical integral in Eq.~\eqref{innerproduct} assuming the LISA 
sensitivity curve, including the contribution 
of the confusion noise from the unresolved Galactic binaries~\cite{Cornish:2018dyw}.
The numerical derivatives of the waveform with respect to the parameters, required to compute the Fisher matrix, are computed as explained in Appendix~\ref{app:derivatives}, whereas the numerical stability of the Fisher and covariance matrices is discussed in Appendix~\ref{app:Fisher}.

We consider $T=1\,{\rm yr}$ observation time, 
with the orbital evolution actually ending not exactly at the $\hat r_{\rm ISCO}$, but at the onset 
of the transition region as defined in~\cite{Ori:2000zn}, i.e. $r_\tn{plunge}=\hat r_{\textup{ISCO}}+\delta \hat r$ 
with $\delta \hat r = 4 q^{2/5}$.
We fix the injected angles to the fiducial 
values $\vartheta_S = \pi/4,\phi_S=0, \vartheta_K=\pi/8, \phi_K =0$. We focus on binaries with 
component masses $M=10^6M_\odot$ and $\mu=10M_\odot$, 
secondary spin $\chi = 0$, setting the primary TLN to 
$k_1 = 0$. Finally, the luminosity distance 
is scaled such that the binary has ${\rm SNR}=30$ for any spin.

\section{Results and discussion}\label{sec:results}
\subsection{Comparison between the fast Fourier transform and the SPA}

We have checked the validity of the SPA by computing the 
\textit{faithfulness} between EMRI waveforms in the 
frequency domain obtained in two differnt ways: (i) with a fast Fourier transform~(FFT) of the time signal~\eqref{eq:gwsignal2}, and (ii) 
with the SPA presented above. Specifically, we compute 
\begin{equation}
\mathscr{F}(h^{\rm{SPA}}_\alpha,h^{\rm{FFT}}_\alpha) = \underset{t_0, \phi_0}{\max}\frac{(h^{\rm{SPA}}_\alpha|h^{\rm{FFT}}_\alpha)}{\sqrt{(h^{\rm{SPA}}_\alpha|h^{\rm{SPA}}_\alpha)(h^{\rm{FFT}}|h^{\rm{FFT}})  }}\ . \label{eq:faithfulness}
\end{equation}

Following the Shannon theorem, for the FFT we use a sampling 
time $\Delta t_s = \lfloor 1/(2 f_{\rm max} ) -1 \rfloor$, 
with $n_s =T/\Delta t_s$ being the total 
number of samples, $T=1\tn{yr}$, and $f_\tn{max}$ given 
by Eq.~\eqref{deffmax}.
Before applying the FFT\footnote{After the tapering we 
have also padded the waveform with $2^{n}$ zeros in order 
to boost the computational speed of the FFT.} we have tapered the 
time domain signal to reduce spectral leakage, using 
a Tukey window with window size $\beta = 0.001$. 

Table~\ref{tab:faifthfullness} provides the values of the 
faithfulness computed for two configurations of the primary 
spin, and for each LISA channel\footnote{
The numerical computation of Eq.~\eqref{eq:faithfulness} can 
be sensitive to the precision adopted in the scalar product. 
For instance, the fractional difference between $\mathscr{F}(h^{\rm{SPA}}_\alpha,h^{\rm{FFT}}_\alpha)$ 
obtained assuming machine precision and 
$40-$digits is at the level of $\mathcal O(1)\%$. 
We have checked the stability of the faithfulness under 
round-off errors by increasing the precision adopted 
in our calculations, finding no changes in the results.}. In agreement with~\cite{Hughes:2021exa}, our results show that the SPA waveform model matches well with the FFT waveform:
$\mathscr{F}(h^{\rm{SPA}}_\alpha,h^{\rm{FFT}}_\alpha)\gtrsim0.993$ 
even for a highly-spinning primary with $\hat{a}=0.99$.
As a useful rule of thumb, values
of $\mathscr{F}$ smaller than $\mathscr{F}\sim1- \frac{{\cal D}}{{2\rm SNR}^2}$, with 
${\cal D}$ dimension of the waveform model, highlights that 
two templates differ significantly among each other~\cite{Lindblom:2008cm,Hughes:2018qxz}. 
For ${\cal D}=12$ as in our case, this threshold translates into $\mathscr{F}\sim 0.993$ 
for SNR=30, so the SPA is sufficiently accurate for a typical EMRI SNR.

As a further assessment of the validity of the SPA, we have compared the standard deviations 
\eqref{fishererrors}, 
obtained with the SPA and 
with the frequency-domain waveforms computed through a FFT.  
In the last case derivatives of the template with respect to 
the binary parameters have been numerically determined 
using a five-point stencil formula (see Ref.~\cite{Piovano:2021iwv} for details), 
except for the luminosity distance $D$, since $\partial \tilde{h}(f)/\partial D$ can be computed analytically.
In the worst case scenario we find that the 
maximum relative difference between the standard deviations 
provided by the Fisher matrix are: (i) $\sim 15\%$ when $\chi$ is 
unbounded, (ii) $\sim 3\%$ when the secondary 
spin is excluded, (iii) $\sim 2\%$ when a prior on 
$\chi$ is imposed.
Overall, these results confirm that the SPA provides 
a reliable and accurate analytic approximation of 
the purely numerically frequency-domain waveforms 
employed for EMRIs.

\begin{table}[ht]
\centering
\begin{tabular}{ccc}
\hline
\hline
$\hat a$ & channel  & $\mathscr F$\\
\hline 
\multirow{2}{*}{0.9} & I   & 0.9931\\   
                     & II  & 0.9970\\ 
\hline
\multirow{2}{*}{0.99} & I & 0.9942\\   
                      & II & 0.9971\\ 
\hline
\hline
\end{tabular}
\caption{Faithfulness $\mathscr{F}(h^{\rm{SPA}}_\alpha,h^{\rm{FFT}}_\alpha)$ 
between frequency-domain waveforms obtained with the SPA, and by applying the FFT to the time domain signal, for different values of the primary spin parameter $\hat{a}$ and for our reference binary system.
The second column identifies the independent channels of the LISA interferometer. We consider EMRIs evolving for one-year up to the plunge. }
\label{tab:faifthfullness}
\end{table}

\subsection{Measurability of the TLN}\label{sec:results_measureTLN}
We now present our main results for the measurability of  the primary TLN, $k_1$. 
In Table~\ref{tab:KerrTeu_5_6PN}, we provide the statistical errors on the waveform parameters when both the 5PN and the 6PN tidal corrections are included in the waveform.\footnote{As shown in Ref.~\cite{Piovano:2021iwv}, the errors on the intrinsic parameters are not significantly affected by the quadrupolar approximation of the waveform. However, including higher multipole moments improves the errors on the luminosity distance $D$ and on the angle $\Delta\Omega_K$, which are therefore overestimated in Table~\ref{tab:KerrTeu_5_6PN}.} We first notice that $k_1$ can be detected with high accuracy even when the secondary spin $\chi$ is considered as an unbounded parameter (first row of Table~\ref{tab:KerrTeu_5_6PN}). The marginalization of $\chi$ (second row) or the inclusion of a Gaussian prior on $\chi$ (third row) improves the statistical error on all intrinsic parameters ($\ln M, \ln \mu, \hat a$, and $k_1$).

We have also checked that including only the leading PN order (5PN) tidal term does not affect significantly the standard deviations on the parameters. This fact provides a good consistency check of our hybrid waveform (mixing BH perturbation theory with PN corrections), since the PN series is not supposed to converge near the ISCO of a highly-spinning BH. We can therefore expect that higher-order tidal terms (or a resummation thereof) would not change our results significantly.

We find that the TLN of the primary $k_1$ can be constrained with the astonishing accuracy of above~$3 \times 10^{-2}$ and $8 \times 10^{-4}$ for $\hat a = 0.9$ and $\hat a  = 0.99$, respectively.
As a figure of merit, it is interesting to note that so far the only measurement of the tidal deformability of a compact object is that coming from GW170817~\cite{LIGOScientific:2018cki}, which set a constraint on the TLN of a neutron star at the level of $\sigma_{k_1}\lesssim 10^3$, i.e. several orders of magnitude less stringent than what achievable with EMRIs.
It is also interesting to note that, for all models of compact objects in which the TLNs scale logarithmically with the compactness~\cite{Cardoso:2017cfl}, $\sigma_{k_1}\sim 10^{-3}$ would allow to probe putative structure at Planckian distance from the horizon and to distinguish between different proposals of exotic compact objects motivated by quantum gravity~\cite{Maselli:2017cmm,Addazi:2018uhd,Maselli:2018fay}.

\begin{figure}[th]
\includegraphics[width=0.45\textwidth]{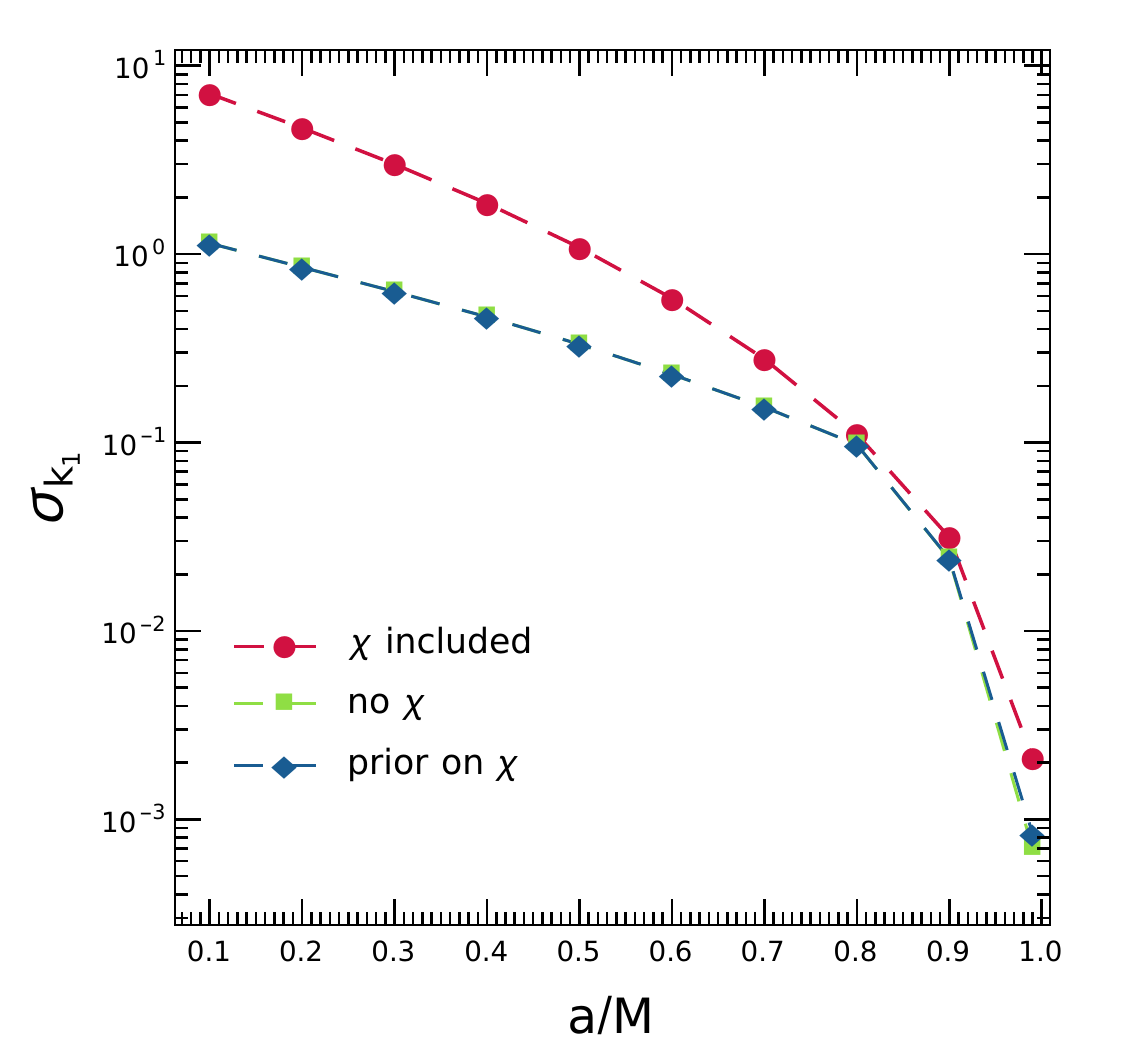}
\caption{1-$\sigma$ error on the primary TLN $k_1$ 
as a function of the primary spin parameter $\hat a\in[0.1,0.99]$. 
Different markers correspond to including the spin of 
the secondary, removing it from the Fisher matrix, 
or adding a Gaussian prior centered around the injected
value of $\chi$ (see main text). The luminosity distance is fixed such that ${\rm SNR}=30$ after 1 year
of observation. 
As a reference, current measurement errors on the TLN of a neutron star coming from GW170817~\cite{LIGOScientific:2018cki} are $\sigma_{k_1}\lesssim 10^3$.
} 
\label{fig:ErrorLovenumber_6PN} 
\end{figure}

Finally, Fig.\ref{fig:ErrorLovenumber_6PN} shows the statistical error on $k_1$ for different values of the primary spin $\hat a$ and different assumptions for the secondary spin.
We observe that the error decreases very rapidly as the primary spin increases, because the majority of the signal comes from the ISCO of the highly-spinning primary, where tidal effects are more relevant. As stressed above, since the PN series poorly converges near the ISCO, it is important that the results shown in Fig.\ref{fig:ErrorLovenumber_6PN} (which includes the 6PN tidal corrections) are very similar to those one would obtain by including only the 5PN terms.
Furthermore, Fig.\ref{fig:ErrorLovenumber_6PN} confirms that neglecting the secondary spin or including it with a very conservative prior gives almost identical results. Thus, the secondary spin does not hamper the ability of measuring a nonstandard small effect such as the primary tidal deformability.
\begin{table*}[th]
\centering
\begin{tabular}{cccccccc|ccc}
\multicolumn{11}{c}{5PN and 6PN TLN terms  \hspace{0.5em} $\hat a_{\text{injected}} = 0.9$}\\
\hline
\hline
prior&$\ln M$  & $\ln \mu$ & $\hat a$ & $\chi $ &$\hat t_0$ &$k_1$ & $\phi_0$ & $\ln D$ & $\Delta \Omega_S$ & $\Delta \Omega_K$\\
  \hline
no &-4.9  & -4.1 & -3.8 & 1.6 & 0.48 & -1.5 & 0.74 &-0.069 &$6.2\times 10^{-4}$ & 7.5 \\
no &-5.8  &-4.2 &  -4.1 &  -- & 0.48  & -1.6 & 0.74 & -0.069 &$5.9\times 10^{-4}$ & 2.9 \\
yes &-5.7  &-4.2 &  -4.1 &  0.57 & 0.48  & -1.6 & 0.74 & -0.069 &$5.9\times 10^{-4}$ & 7.5 \\
  \hline
  \hline
  \multicolumn{11}{c}{\vspace{0.3em}} \\
  \multicolumn{11}{c}{5PN and 6PN TLN terms  \hspace{0.5em} $\hat a_{\text{injected}} = 0.99$}\\
 \hline
 \hline
prior &$\ln M$  & $\ln \mu$ & $\hat a$ & $\chi $ &$\hat t_0$ &$k_1$ & $\phi_0$ & $\ln D$ & $\Delta \Omega_S$ & $\Delta \Omega_K$\\
  \hline
no &  -5.2  & -4.6 & -4.4 & 1.2 & 0.21 & -2.7 & 0.74 &-0.071 &$2.7\times 10^{-4}$ & 6.7 \\
no & -5.8  &-4.9 &  -5.0 &  -- & 0.21  & -3.1 & 0.74 & -0.071 &$2.7\times 10^{-4}$ & 2.6 \\
yes & -5.7  &-4.8 &  -4.9 &  0.61 & 0.21  & -3.1 & 0.74 & -0.071 &$2.7\times 10^{-4}$ & 6.7 \\
  \hline
  \hline
\end{tabular}
\caption{Top: Fisher-matrix errors on the intrinsic source parameters, on the luminosity distance, and on the solid angles which define the orientation and the orbital angular momentum of the binary for our model. The primary has spin $\hat{a}=0.9$ while the secondary is nonspinning, with $M = 10^6 M_\odot$, $\mu = 10 M_\odot$, and $k_1 = 0$. 
For clarity, we present the $\log_{10}$ of the errors on $\ln M$, $\ln \mu$, $\hat a$, $\chi$, $\hat t_0$, $k_1$, $\phi_0$, and $\ln D$.  We include both the 5PN and 6PN TLN terms.
Bottom: same as the top but with $\hat a = 0.99$. 
The SNR for a source at $D=1\,{\rm Gpc}$ is ${\rm SNR}=111$ (top) and ${\rm SNR}=125$ (bottom), but the errors are all normalized to the fiducial value ${\rm SNR}=30$. }\label{tab:KerrTeu_5_6PN}
\end{table*}
%

\section{Conclusion}\label{sec:conclusion}
Measuring a nonzero TLN for a supermassive object would be a robust smoking gun for new physics beyond the standard BH prediction in GR.
EMRIs detectable by LISA are unique sources for tests of gravity and allow for unparalleled measurements of beyond-GR effects. 
With these motivations in mind, we have estimated the accuracy in the measurement of the tidal deformability of a supermassive compact object through an EMRI detection by LISA.
Confirming back-of-the-envelope estimates~\cite{Pani:2019cyc}, we found the TLN of the central supermassive object can be measured at the level of $10^{-3}$ if the central object is highly spinning. This is about 6 orders of magnitude better than current accuracy in measuring the TLNs of a NS with ground-based detectors.

We included the secondary spin as a possible source of confusion, showing that its inclusion does not affect the bounds on the primary TLN. On the other hand, we have focused on simplified (circular, equatorial, nonprecessing) orbits. It would be important to extend our analysis by including eccentricity, inclined orbits~\cite{Hughes:2021exa,Katz:2021yft,Skoupy:2021asz}, and possible spin precession~\cite{Drummond:2022efc,Witzany:2019nml}. On the one hand these extensions will increase the dimensionality of the parameter space, rendering parameter estimation more demanding, but on the other hand they might also help in disentangling possible parameter correlations.
Another possible extension would be the inclusion of important post-adiabatic corrections to the waveforms~\cite{Wardell:2021fyy,Mathews:2021rod,Lynch:2021ogr,Huerta:2008gb}.
Finally, we have adopted a hybrid “Teukolsky+PN” waveform, where tidal corrections were introduced with their corresponding (leading and next-to-leading order) PN terms. An interesting extension would be to compute the tidal deformability contribution in the point-particle limit but without PN expansion, by evaluating the tidal tensor of the secondary along its worldline.

As a by-product of our analysis, we have assessed the accuracy of the SPA to perform efficient tests of gravity with EMRI waveforms in the frequency domain. 
One great advantage of the SPA is that it reduces the number of numerical derivatives required to compute and invert the Fisher matrix, making the error estimate extremely more efficient from a numerical perspective.
Although we have applied this approach to the specific case of constraining the TLNs, we expect the same method can be applied to several other tests of gravity.
In a future work~\cite{PiovanoInPrep} we will use this approach to constrain a comprehensive parametrized waveform accounting for multiple deviations at the same time.

\begin{acknowledgments}
We thank Niels Warburton for interesting discussions about the SPA.
This work makes use of the Black Hole Perturbation Toolkit. 
Numerical computations were performed at the Vera cluster of the Amaldi Research Center funded by the MIUR program  ``Dipartimento di 
Eccellenza" (CUP:~B81I18001170001).
P.P. acknowledge financial support provided under the European Union's H2020 ERC, Starting Grant agreement no.~DarkGRA--757480. We also acknowledge support under the MIUR PRIN and FARE programmes (GW-NEXT, 
CUP:~B84I20000100001), and networking support by the COST Action CA16104.
This work is partially supported by the PRIN Grant 2020KR4KN2 ``String Theory as a bridge between Gauge Theories and Quantum Gravity''.
\end{acknowledgments}

\appendix

\section{Semi-analytic derivatives of the waveforms} \label{app:derivatives}
The frequency-domain waveform~\eqref{eq:gwSPA} has an implicit dependence on the intrinsic parameters $\vec x\equiv \vec{y}_I = (\ln M, \ln \mu, \hat a, \chi, k_1)$ through the functions $\tilde t(f;\vec x)$, $\phi(\tilde t(f;\vec x);\vec x)$ and $\dot \Omega(\tilde t(f;\vec x);\vec x)$. In this appendix 
we show how to compute the derivatives of the 
waveforms with respect to the intrinsic parameters 
$\vec x$ in a semi-analytic fashion.

By the theorem of the implicit functions, the derivatives $\partial \tilde t(f;\vec x)/\partial x^i$ are given as
\begin{equation}
\frac{\partial \tilde t(f;\vec x)}{\partial x^i} = \left.-\frac{1}{\dot \Omega(t;\vec x)} \frac{\partial\Omega(t; \vec x)}{\partial x^i} \right \rvert_{t = \tilde t(f;\vec x)}\ .
\end{equation}
The derivatives $\partial\phi(t;\vec x)/\partial x^i$ are 
instead given as solutions 
of the following ordinary differential equation with initial condition $\partial\phi(0;\vec x)/\partial x^i=0$:
\begin{equation}
\totder{}{t}\bigg(\frac{\partial \phi(t;\vec x)}{\partial x^i}\bigg) = \frac{\partial\Omega(\hat r(t;x); \vec x)}{\partial x^i} + \frac{\partial\Omega(\hat r(t;x); \vec x)}{\partial \hat r}\frac{\partial \hat r(t;\vec x)}{\partial x^i}\ ,
\end{equation}
where $\partial\hat r(t;\vec x)/\partial x^i$ can be 
computed from
\begin{align}
\totder{}{t}\bigg(\frac{\partial \hat r(t;\vec x)}{\partial x^i}\bigg) &= \frac{\partial}{\partial x^i}\bigg(\totder{\hat r}{t}(\hat r(t;\vec x);\vec x)\bigg) + \nonumber\\
&+\frac{\partial}{\partial \hat r}\bigg(\totder{\hat r}{t}(\hat r(t;\vec x);\vec x)\bigg)\frac{\partial \hat r(t;\vec x)}{\partial x^i}
\end{align}
with initial condition $\partial\hat r(0;\vec x)/\partial x^i = 0$.
Finally, the derivatives $\partial \phi(\tilde t(f;\vec x);\vec x)/\partial x^i$ can be written as
\begin{align}
\frac{\partial \phi(\tilde t(f;\vec x);\vec x)}{\partial x^i} &= \left.\frac{\partial \phi(t;\vec x)}{\partial x^i}\right \rvert_{t = \tilde t(f;\vec x)} + \totder{\phi(t;\vec x) }{t} \frac{\partial \tilde t(f;\vec x)}{\partial x^i} \nonumber\\
&= \left.\frac{\partial \phi(t;\vec x)}{\partial x^i}\right \rvert_{t = \tilde t(f;\vec x)} + \pi f\frac{\partial \tilde t(f;\vec x)}{\partial x^i}
\end{align}
Therefore, the derivatives $\partial \Phi_\alpha (\tilde t(f;\vec x))/\partial x^i$ of the SPA 
phase~\eqref{eq:SPAphase} are
\begin{equation}
 \frac{\partial \tilde\Phi_\alpha (\tilde t(f;\vec x);\vec x)}{\partial x^i} = \left.-2 \frac{\partial \phi(t;\vec x)}{\partial x^i}\right \rvert_{t = \tilde t(f;\vec x)}\ .
\end{equation}

Note that the contribution to $\partial\tilde\Phi_\alpha (\tilde t(f;\vec x))/\partial x^i$ given by $\partial \phi^{\text{sh}}_\alpha (\tilde t(f;\vec x))/\partial x^i$ and $\partial \phi^{\text{Dop}}_\alpha (\tilde t(f;\vec x))/\partial x^i$ is negligible, since $\Omega(t) T_{\rm LISA}\gg 2\pi$ for a typical EMRI detectable by LISA. 

Finally, the derivatives of the frequency sweep $\partial \dot \Omega(\tilde t(f;\vec x);\vec x)/\partial x^i$ 
are given by
\begin{equation}
 \frac{\partial \dot \Omega(\tilde t(f;\vec x);\vec x)}{\partial x^i} = \left( \left. \frac{\partial \dot \Omega(t;\vec x)}{\partial x^i} + \ddot \Omega(t;\vec x)\frac{\partial \tilde t(f;\vec x)}{\partial x^i}\right) \right\rvert_{t = \tilde t(f;\vec x)}\ .
\end{equation}
Once $\partial\tilde t(f;\vec x))/\partial x^i$, $\partial \Phi_\alpha (\tilde t(f;\vec x))/\partial x^i$ and  $\partial\dot \Omega(\tilde t(f;\vec x);\vec x)$ are known, 
the semi-analytic derivatives of the frequency 
domain template~\eqref{eq:gwSPA} with respect to 
the binary parameters can be constructed 
straightforwardly.
%
\section{Stability of the Fisher matrix} \label{app:Fisher}

In this appendix we provide further details on the accuracy 
of the calculations we performed, assessing the numerical 
stability of the covariance matrix for the waveform parameters. 
This is particularly relevant in the case of EMRIs, for which 
Fisher matrices are known to be ill-conditioned~\cite{Vallisneri:2007ev}, and 
small numerical or systematic errors are 
amplified after computing the inverse. 
As a rule of thumb, for a condition number 
$\kappa=10^n$, one may lose up to $n$ figures of 
accuracy, which should be added to the numerical 
errors.

This problem is exacerbated when finite-difference 
methods are employed for the waveform derivatives \cite{Piovano:2021iwv,Burke:2020vvk}, 
since the covariance matrix can be sensitive to the 
choice in the parameter shifts adopted for the 
differentiation.
However, the semi-analytic approach described in Appendix~\ref{app:derivatives}, combined with 
the SPA, avoids such issues.

Inverting the Fisher matrices still remains a 
delicate task, which can depend on the numerical precision 
used for the calculation due to the large condition number. 
Indeed, for the binary configurations we considered, 
we find $\kappa \sim 10^{25}$ and $\kappa \sim 10^{18}$, 
for a primary with $\hat a =0.1$ and 
$\hat a =0.99$, respectively.

We have first tested the stability of the Fisher inversion 
against changes in the numerical precision. 
In the worst case, which occurs when 
the secondary spin is included, we find that a stable 
covariance matrix requires at least 35 digits 
of precision in input.

Moreover, we have checked the sensitivity of both 
Fisher and covariance matrices to small variations of their 
components, by perturbing them with a deviation matrix 
$F^{ij}$. We draw all elements of $F^{ij}$ from a 
uniform distribution $U\in[a,b]$, and then compute
\begin{equation}
\delta_{\textit{\rm stability}} \equiv \underset{ij}{\max} \Bigg[ \frac{\big((\Gamma+F)^{-1}-\Gamma^{-1}\big)^{ij}}{(\Gamma^{-1})^{ij}}\Bigg]\ .
\end{equation}

For the most problematic configurations we analysed:

\begin{itemize}
    \item the inverse without priors is stable with $\delta_{\textit{\rm stability}} = 5\%$ with perturbations $U\in[-10^{-9},10^{-9}]$.
     \item the inverse with priors is stable with $\delta_{\textit{\rm stability}} = 6\%$ with perturbations $U\in[-10^{-6},10^{-6}]$.
     \item the inverse without secondary spin $\chi$ is stable with $\delta_{\textit{\rm stability}} = 2\%$ with perturbations $U\in [-10^{-5},10^{-5}]$.
\end{itemize}

The stability of the Fisher matrices drastically 
improves as the spin of the primary increases. 
For $\hat a = 0.99$, the inverse is stable with 
$\delta_\tn{stability} \lesssim2\%$ and 
perturbations $U\in [-10^{-5},10^{-5}]$ for 
all cases we considered.



\bibliographystyle{utphys}
\bibliography{Ref}

\end{document}